\documentclass[aps,pra,twocolumn,superscriptaddress,amsmath]{revtex4-1}

\pdfoutput = 1

\usepackage[pdftex]{graphicx,color}

\usepackage{mathtools}
\usepackage{amsfonts}
\usepackage{amssymb}
\usepackage{amsmath}
\usepackage{amsthm}
\usepackage{times}

\usepackage{subfigure}
\usepackage{rotate}
\usepackage{bm}
\usepackage{dcolumn}

\usepackage{hyperref}

\let\oldmarginpar\marginpar
\renewcommand\marginpar[1]{\-\oldmarginpar[\raggedleft\tiny #1]%
{\raggedright\tiny #1}}

\newcommand{\bra}[1]{\langle#1|}
\newcommand{\ket}[1]{|#1\rangle}

\newcommand{\up}{\uparrow}
\newcommand{\down}{\downarrow}

\newcommand{\f}[2]{{\ensuremath{\mathchoice%
       {\dfrac{#1}{#2}}
       {\dfrac{#1}{#2}}
       {\frac{#1}{#2}}
       {\frac{#1}{#2}}
       }}}

\newcommand{\set}[1]{\{ #1 \}}
					
\newcommand{\Set}[1]{\left\{ #1 \right\}}

\newcommand{\Sc}{S_{\mathrm{core}}}

\begin{document}


\title{Clustering in Hilbert space of a quantum optimization problem}
\hypersetup{pdftitle={Quantum clustering at criticality in QSAT},
	pdfauthor={C. R. Laumann}}
	

\author{S. C. Morampudi}
\affiliation{Department of Physics, Boston University, Boston, MA 02215, USA}
\affiliation{Max Planck Institut f¨ur Physik komplexer Systeme, %
	01187 Dresden, Germany}

\author{B. Hsu}
\affiliation{Department of Physics, Princeton University, Princeton, NJ 08544, USA}

\author{S. L. Sondhi}
\affiliation{Department of Physics, Princeton University, Princeton, NJ 08544, USA}

\author{R. Moessner}
\affiliation{Max Planck Institut f¨ur Physik komplexer Systeme, %
	01187 Dresden, Germany}

\author{C. R. Laumann}
\affiliation{Department of Physics, Boston University, Boston, MA 02215, USA}

\date{\today}


\begin{abstract}
The solution space of many classical optimization problems breaks up into  clusters which are extensively distant from one another in the Hamming metric.
Here, we show that an analogous quantum clustering phenomenon takes place in the ground state subspace of a certain quantum optimization problem.
This involves extending the notion of clustering to Hilbert space, where the classical Hamming distance is not immediately useful. 
Quantum clusters correspond to macroscopically distinct subspaces of the full quantum ground state space which grow with the system size.
We explicitly demonstrate that such clusters arise in the solution space of random quantum satisfiability (3-QSAT) at its satisfiability transition.
We estimate both the number of these clusters and their internal entropy. 
The former are given by the number of hardcore dimer
coverings of the core of the interaction graph, while the latter is related to the underconstrained degrees of freedom not touched by the dimers.
We additionally provide new numerical evidence suggesting that the 3-QSAT satisfiability transition may coincide 
with the product satisfiability transition, which would imply the absence of an intermediate entangled satisfiable phase.
\end{abstract}

\maketitle


\section{Introduction} 
\label{sec:introduction}

The concept of a `rugged energy landscape' plays a central role in many parts of statistical physics and optimization theory.
The idea is that in glassy systems, this energy landscape in configuration space becomes labyrinthine and difficult to explore with local dynamics. 
Optimal configurations disappear into deep, narrow valleys which are hard to find amidst the hills and numerous shallow valleys. 
Classically, this geometric intuition has been at least partially quantified by studying the statistical `clustering' of low energy configurations into distinct locally connected regions separated by a macroscopic Hamming distance (which measures distance between two solution strings by looking at the number of characters which differ).
Such clustering has been used to understand a large range of classical phenomena including the slow down of combinatorial optimization algorithms in certain regimes~\cite{Mezard2005, Mulet2002, Cocco2003, Martin2004}, dynamical freezing in spin glasses~\cite{SpinGlassTheoryBook1986}, stable associative memories in neural networks~\cite{Hopfield1982}, and even improved error correction algorithms in classical codes~\cite{Montanari2001}.

Although clustering has been argued to play a role in quantum mechanical systems (see below), it is less clear how to construct an intrinsic geometric understanding of clustering in Hilbert space.
There are several essential difficulties. 
A low energy quantum mechanical state space is a continuous linear vector space where the simplest notion of distance is the Fubini-Study metric which measures the overlap between two states. However, this ascribes the same distance to states which differ by a single spin flip as to those which are macroscopically rearranged. 
This clearly fails to capture the much larger distance between the states according to local dynamics and local physical expectation values.
On the other hand, a purely discrete distance such as the Hamming weight defined by the minimal support of operators mapping between two states is far too sensitive to small continuous deformations in Hilbert space. 
Moreover, quantum mechanics allows superpositions of macroscopically distinct states. 
How should one measure the distances between such `cat'-states?
Finally, one would like to extend these notions to allow `local degeneracy' within a cluster.


To overcome these difficulties, we propose to focus on short-range correlated states %
\footnote{Such states have been called `pure' or `clustering' states in the literature, but we will avoid this terminology to prevent confusion with the quantum mechanical definition of pure states and the notion of clustering defined in solution space.} 
whose connected correlations decay at large physical separation~\cite{SpinGlassTheoryBook1986}. That is,
\begin{align}
	\bra{\psi} O_i O_j \ket{\psi} \xrightarrow[|i-j| \to \infty]{} \bra{\psi}O_i\ket{\psi}\bra{\psi} O_j \ket{\psi}
\end{align}
for local operators $O_i$ and $O_j$ centered on sites $i$ and $j$, where $|i - j|$ measures the interaction graph distance between $i$ and $j$. 
Such states are as far as possible from `cat' states (macroscopic superpositions). 
In order for a low energy subspace to be clustered, we take as prerequisite that it must contain a large number of such short-range correlated states which are macroscopically distinct, i.e. for which all local operators have vanishing off-diagonal matrix elements when the number of qubits $N\rightarrow\infty$ (which we refer to as the thermodynamic limit in the following).
This definition is analogous to that used to identify distinct symmetry breaking vacua (such as in ferromagnets), but here we do not expect that the short-range correlated states be related by symmetries. 

This set of reference states may not provide a complete or unique basis for the low energy subspace, especially if there are local degeneracies. 
Rather, it is natural to consider the state space clustered if it can be decomposed into a sum of macroscopically distinct subspaces (which become orthogonal in the thermodynamic limit as we describe below), and each of which can be constructed from local operations acting on a reference short-range correlated state.

A prominent setting where classical clustering arises is in random ensembles of the canonical NP-complete problem of satisfiability ($k$-SAT).
Satisfiability consists of determining whether a given $k$-body classical Hamiltonian composed of a sum of classical projectors has zero-energy (`satisfying') states~\cite{Mezard2005}. 
Quantum satisfiability ($k$-QSAT) generalizes classical $k$-SAT by allowing the terms to be general non-commuting projectors~\cite{Bravyi2006}.
Random $k$-QSAT thus provides a natural place to look for quantum clustering.
Random $k$-QSAT is known to exhibit both satisfiable (SAT) and unsatisfiable (UNSAT) phases as in classical satisfiability.
However, the phases show finer structure due to the possibility of quantum entanglement. 
More precisely, the SAT phase in general contains two distinct phases - one where the ground states include product states (PRODSAT) \cite{LaumannQSAT2010,LaumannQSAT2010_2} and another where the only ground states present are entangled (ENTSAT) \cite{Ambainis2012}. 
For a more detailed review of this phase diagram, see \cite{Sattath2016,Hsu2013,Bravyi:2009p7817}.

\begin{figure}
 	\includegraphics[width=1.0\columnwidth]{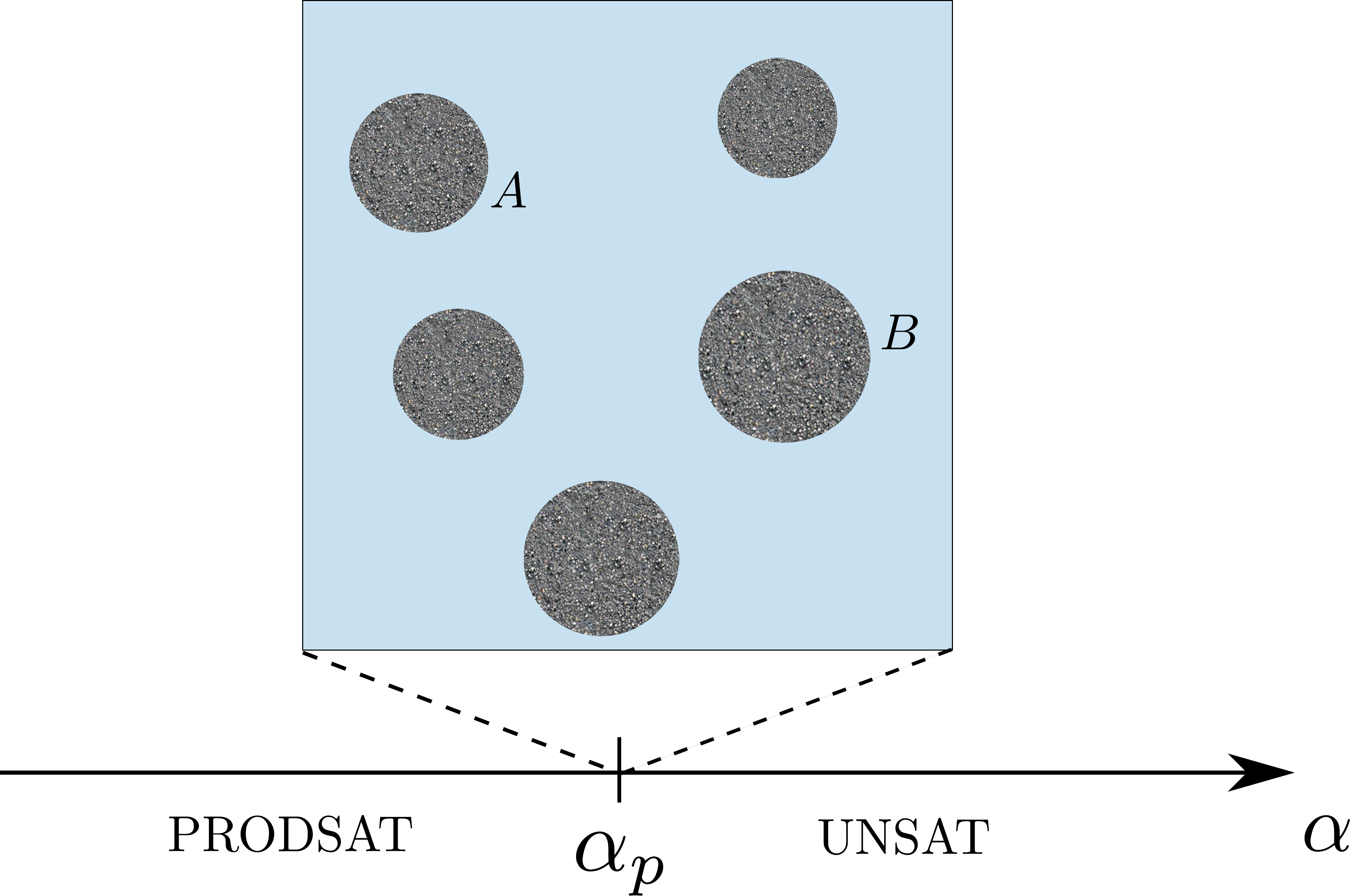}
	\caption{
    \label{fig:clusteringSchematic} 
    Illustration of clusters of solutions (grey discs) in Hilbert space at the product satisfiability transition $\alpha_p$ in 3-QSAT.
    Random quantum satisfiability exhibits a satisfiable phase with non-entangled satisfying states  (PRODSAT) at small clause density $\alpha$ and an unsatisfiable (UNSAT) phase at large clause density. For $k$ sufficiently large, these are separated by a satisfiable phase with only entangled satisfying states, but for $k=3$, the numerics in Sec.~\ref{sec:new_numerical_bounds_on_unsat_transition_for_} suggest that the transition from PRODSAT to UNSAT is direct at $\alpha_{p} \approx 0.91$. 
    }
\end{figure}

In this paper, we show that quantum clustering arises in the zero-energy space of 3-QSAT at the critical point out of the PRODSAT phase. 
Our argument is constructive, as we associate the cluster subspaces with discrete product state reference states on an appropriately defined core of the critical instances. 
These reference states are further in correspondence with dimer coverings of the core, which we show are both exponentially numerous and macroscopically distinct - leading to the same properties for the reference states. 
Finally, each reference state extends into its own exponentially large satisfying subspace when additional degeneracy associated with the non-core qubits (the `hair') is taken into account.


In the bulk of the paper, we make these arguments more precise.
In Sec.~\ref{sec:QSAT}, we review QSAT and its various phases, and explain the mapping of product ground states to dimer coverings. 
We then show how this leads to quantum clustering in the 3-QSAT ground space at the PRODSAT critical point.
In Sec.~\ref{sec:new_numerical_bounds_on_unsat_transition_for_}, we provide numerical evidence suggesting that there is a direct transition to an UNSAT phase from the PRODSAT phase in 3-QSAT -- that is, that the ENTSAT phase does not exist at $k=3$. 
This helps to motivate our focus on the PRODSAT transition, and is also of independent interest in pinning down the phase diagram of random $k$-QSAT.
In Sec.~\ref{sec:counting_dimer_covering_states}, we compute the entropy of clusters $\Sc$ (\emph{ie.} the logarithm of the number of cluster subspaces) using cavity techniques on the associated dimer covering problem.
In Sec~\ref{sub:relation_to_qsat}, we provide an estimate of the internal entropy of a cluster $S_{\mathrm{hair}}$ by considering the contribution of the non-core degeneracy.
Finally, we conclude briefly in Sec.~\ref{sec:conclusion}.


There have been several streams of previous work on clustering in random quantum mechanical systems.
The most direct connections are provided by introducing weak quantum terms in classical models.
The clustering properties of the classical energy landscape then provide the backdrop for the tunneling induced by the quantum terms, with consequences for quantum spin glass dynamics~\cite{Goldschimdt1990, Laumann2014,Baldwin:2017aa, Laumann2008} and the adiabatic algorithm~\cite{Knysh:2006p9709,Knysh:2008dk,Altshuler2010, Foini2010, Bapst2013}. 
These treatments take the classical clustering properties as a starting point and do not try to extend them to an intrinsic notion of clustering with finite transverse field.
On the other hand, the replica treatment of certain mean-field spin glass models exhibits replica symmetry breaking in imaginary time even at finite fields~\cite{Goldschimdt1990, Ye1993}.
In the classical case, replica symmetry breaking is often identified with clustering in configuration space, so this provides a possible approach to identifying clustering in quantum models.
However, this microscopic interpretation of replica symmetry breaking is not self-evident, especially in the quantum case.
Finally, the solvable AKLT spin glass model lifts the classical clustering exhibited by a finite temperature vector spin glass into a quantum mechanically degenerate set of macroscopically distinct ground states~\cite{Laumann2010}. 
This work bears the closest resemblance to the structure that we report in this paper.

Several recent works have explored the computational difficulty of identifying whether two given low energy states of a local Hamiltonian can be connected by a sequence of $k$-local operations which remain at low energy~\cite{Gharibian2015, Gosset2016}. 
The essential result is that determining whether such low energy paths exists is quite difficult in the worst case -- QCMA-complete, to be precise. 
While this notion of path traversal is closely related to clustering, it requires more detailed control of the excitation energetics than we have in random QSAT. 
It would be very interesting to develop a statistical picture of such paths in this model.

\begin{figure}
 	\includegraphics[width=1.0\columnwidth]{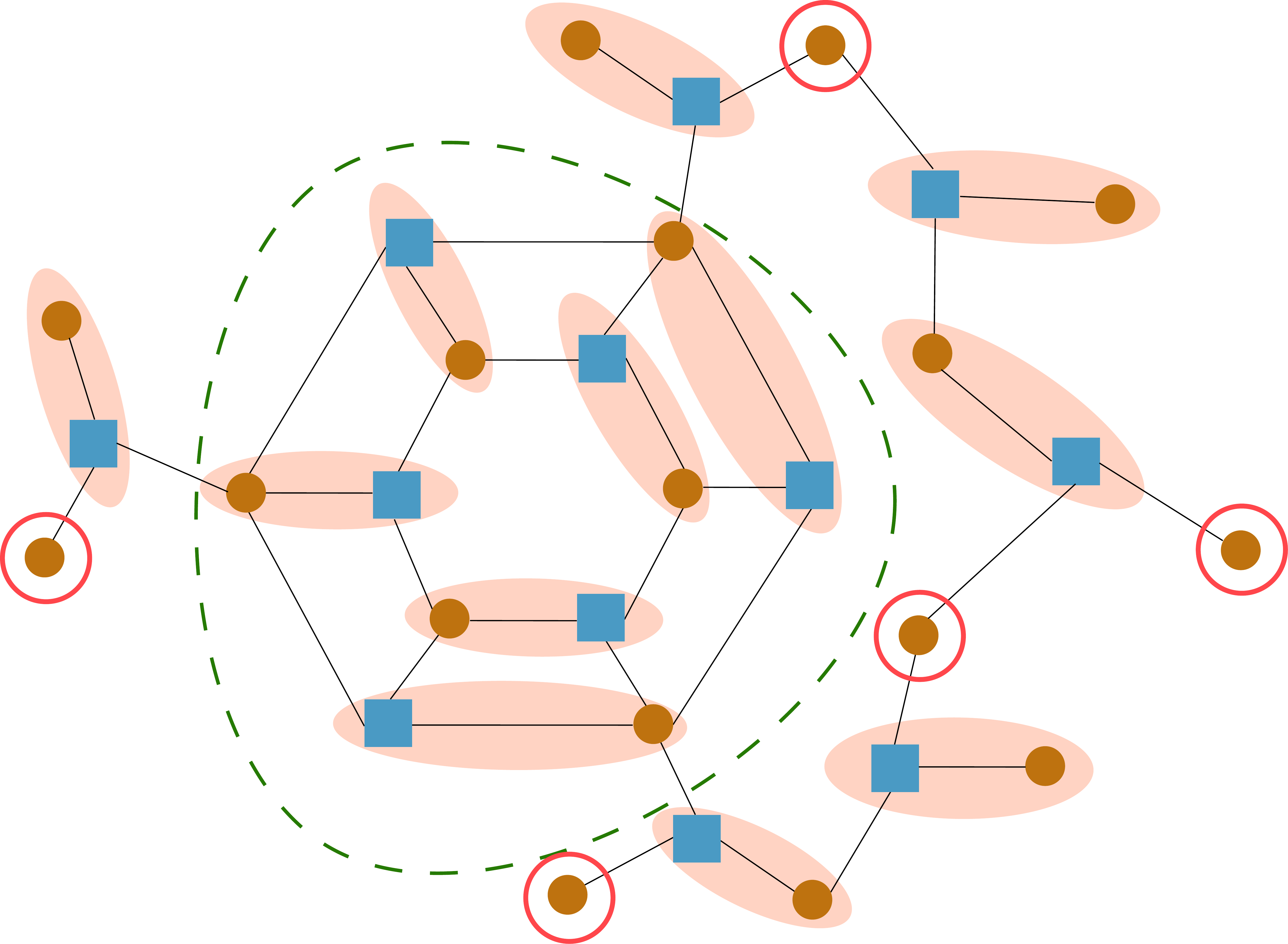}
	\caption{\label{fig:interactionGraph} 
    Interaction graph representation of an instance of $3$-QSAT. 
    Projectors (blue squares) act on the 3 qubits (brown circles) to which they are connected.
    In a product satisfiable instance, dimers (salmon ovals) cover every projector but may leave some qubits free (monomers, circled)
    On a critical core (inside green dashed loop), however, a dimer covering leaves no qubit free. The only free qubits (circled) lie on the hair of the graph (outside green dashed loop).
    }
\end{figure}


\section{Technical Background}\label{sec:QSAT}

The rank 1, $k$-QSAT Hamiltonian on $N$ qubits is given by 
\begin{align}
H = \sum_{m=1}^{M} \Pi_m = \sum_{m=1}^{M} \left | \phi_m \rangle \langle \phi_m \right | 
\end{align}
where each \emph{clause}, $\Pi_m$, is a rank 1 projector acting on a group of $k$ qubits labeled by $m = (m_1, m_2, \cdots, m_k)$. 
Pictorially, we represent $H$ graphically through an interaction graph $G$ where the clauses are represented by squares and the qubits by circles, as in Fig.~\ref{fig:interactionGraph}.
Each clause $\Pi_m$ projects onto one state $\ket{\phi_m}$ in the local $2^k$-dimensional Hilbert space, imposing an energy cost on any state which overlaps that direction. 
Such a Hamiltonian has a zero energy state $\ket{\Psi}$ if and only if $\ket{\Psi}$ is simultaneously annihilated by all of the projectors $\Pi_m$. 
In this case, we say that $H$ is \emph{satisfied} by the state $\ket{\Psi}$. 
Deciding whether a given $k$-QSAT Hamiltonian is satisfiable is QMA$_1$-complete for $k \ge 3$~\cite{Bravyi2006, Gosset2013}, modulo certain technical restrictions. 

We focus on the prototypical random $k$-QSAT ensemble introduced in \cite{LaumannQSAT2010_2}.
There are two sources of randomness: 
(1) the discrete choice of interaction graph $G$ and
(2) the continuous choice of projectors $\Pi^m$ associated to each interaction.
The latter choice is particularly powerful: \emph{generic} choices\footnote{We use the term \emph{generic} to refer to the continuous choice of projectors and \emph{random} to refer to the choice of graph $G$.} of projectors reduce quantum satisfiability to a graph, rather than Hamiltonian, property \cite{LaumannQSAT2010_2}. 
More precisely, for fixed $G$, the dimension $R_G = |\ker H|$ of the satisfying subspace is \emph{almost always} minimal with respect to the continuous choice of $\Pi^m \in \mathbb{CP}^{2^k-1}$. 
We refer to this as the ``geometrization'' property.
The satisfying dimension $R_G$ may be lower bounded using the Quantum Shearer formula \cite{Sattath2016} and in certain limits it is conjectured that this formula gives a tight, and thus generic, result.

For generic QSAT instances, zero energy \emph{product} states exist if and only if the graph $G$ admits ``dimer coverings'', which match qubits (circles) to interactions (squares) covering all of the interactions (see Fig.~\ref{fig:interactionGraph}) \cite{LaumannQSAT2010}. When the graph is fully packed, every interaction is covered by exactly one dimer. Away from full packing, there are uncovered qubits or interactions which are ``monomers'' on the interaction graph.
The correspondence between product ground states and dimer coverings is clearest in the case of product projectors $\Pi^m = \Pi^m_1 \otimes \cdots \otimes \Pi^m_k$. 
Such projectors are satisfied if and only if one of the single qubit factors $\Pi^m_i$ is satisfied, which in turn specifies that the qubit on which $\Pi^m_i$ acts is in a local state orthogonal to $\Pi^m_i$.
A dimer covering provides a matching between projectors $m$ and the choice of qubit that satisfies $m$.
More detailed arguments are required to show that these satisfying product states may be extended to generic (non-product) projectors \cite{LaumannQSAT2010}. The upshot is that there is a mapping from dimer coverings $DC$ of the interaction graph to product states $\ket{DC}$.

We follow previous work and choose the random interaction graph $G$ according to an Erd{\"o}s-Renyi ensemble with clause density $\alpha = M/N$.
That is, each of the possible $\binom{N}{k}$ interactions is placed with independent probability $p = M / \binom{N}{k}$. In this ensemble, each clause has exactly $k$ neighboring qudits while qudit $i$ has $d_i$ neighboring squares where $d_i$ a random degree which is Poisson distributed with mean $\overline{d_i} = k \alpha$.
The clause density $\alpha$ provides a tuning parameter to explore $k$-QSAT instances ranging from extremely underconstrained at small $\alpha$ to severely overconstrained at large $\alpha$. 

One of the most important geometric features of the Erd\"os-Renyi interactions graphs is the existence of a core above a critical $\alpha_{hc}$.  
The core $G' \subset G$ is the maximal subgraph of $G$ such that every qubit has degree at least two %
\footnote{In much of the relevant literature, this core is refered to as a hypercore as, strictly speaking, $G'$ is the maximal sub-hypergraph of the hypergraph $G$ where the interactions are viewed as hyperedges. In this paper, we eschew this unnecessarily hyperactive terminology.}.
The core may be constructed efficiently from a given graph $G$ by recursively removing the `dangling' degree 1 qubits along with the neighboring interactions until no degree 1 qubits remain \footnote{We also remove any isolated degree 0 qubits.}. 
We refer to the part of the graph removed by this process (\emph{i.e.} $G-G'$) as the \emph{hair} of $G$ (see Fig.~\ref{fig:interactionGraph}).

\section{Quantum clustering at the PRODSAT transition}
In this section, we show the ground space of random QSAT is clustered at the critical point $\alpha_{p}$ out of the PRODSAT phase. 

To recap, for quantum clustering, we seek a set of short-range correlated reference states each of which generate macroscopically distinct subspaces through local operations at zero energy. More specifically, we need 

(i) a set of reference states whose connected correlators of local operators vanish and from which subspaces can be generated through local operations. In QSAT, these reference states correspond to product states as described below.

(ii) local operators should have vanishing off-diagonal matrix elements between states in different subspaces 

We construct the reference states in two steps. 
First, we choose a product state $\ket{DC}$ on the core of the interaction graph. 
These are in one-to-one correspondence with dimer coverings (DC) of the core.
Second, we extend this dimer covering to the full interaction graph and correspondingly extend the reference state to a product state on the full graph.
The extension is not unique, even among product states, but this non-uniqueness corresponds to local degeneracy.
It is precisely this degeneracy which generates the subspace $K^{DC}_{\mathrm{hair}}$ associated to each reference state.
This leads to a clustering decomposition of the full ground space $K$ as
\begin{equation}
K=\sum_{DC}|DC\rangle_{\mathrm{core}} \bigotimes K^{\mathrm{DC}}_{\mathrm{hair}}
\label{eq:DCHair}
\end{equation}

Since the reference states are product states, their connected correlators vanish, i.e., $\langle \hat{O}_i \hat{O}_j \rangle \rightarrow \langle \hat{O}_i \rangle \langle \hat{O}_j \rangle$ whenever $\hat{O}_i$ and $\hat{O})_j$ have non-intersecting support, and certainly when $|i - j| \to \infty$. These then satisfy the first condition for clustering. 

If we consider the situation when the dimer covering is fully packed on the core ($M_c=N_c$, i.e, $\beta=\beta_c=1$), the only way to go from one dimer configuration on the core to another is to re-arrange along a loop on the graph due to the absence of monomers (free qubits). It is important to note that the presence of monomers on the hair (Fig.~\ref{fig:interactionGraph}) does not allow us to rearrange the fully packed dimer configuration on the core. Then the typical overlap between two states $|DC_1 \rangle$ and $|DC_2 \rangle$ is 
\begin{equation}
\overline{\log\langle DC_1 | DC_2 \rangle \strut} \approx -\gamma \mathcal{L}[DC_1, DC_2]
\end{equation}
where $\mathcal{L}[DC_1, DC_2]$ is the length of the loops formed by superposing the two dimer coverings onto each other, and $\gamma$ is related to the $O(1)$ overlap of two random vectors on the Bloch sphere. Now if the length of the loops increase with system size, then the overlap vanishes in the thermodynamic limit and the states are orthogonal. More importantly, the expectation value of local operators also similarly decay and  have vanishing off-diagonal components between the states in the thermodynamic limit.

So the question is whether there is a regime in 3-QSAT where we only have large loops. In fact, this occurs at the transition point out of the PRODSAT phase. Although there is a finite extent of clause density where we have only large loops, it is only at the transition point that we have fully packed dimer coverings of the core. These cannot be rearranged to each other through local changes such as hopping unpaired spins (monomers) to neighboring positions. We show in Sec.~\ref{sec:counting_dimer_covering_states} that there is an extensive number of dimer coverings on the core ($|DC\rangle_{\mathrm{core}}$) with a corresponding entropy ($\Sc$) at the transition point. When we add the hair back to the core, it generates a continuous manifold of states around each dimer covering, thus enlarging each point to a cluster ($K^{\mathrm{DC}}_{\mathrm{hair}}$). We estimate the internal entropy of this cluster ($S_{\mathrm{hair}}$) in Sec.~\ref{sub:relation_to_qsat}. 

\section{New Numerical Bounds on UNSAT Transition for $k=3$} 
\label{sec:new_numerical_bounds_on_unsat_transition_for_}

The question of the properties of the endpoint of the PRODSAT phase goes hand in hand with that about the nature of the adjacent phase. 
The two generic possibilities are that there is either an (entangled) satisfiable phase, ENTSAT, as happens for $k$-QSAT for sufficiently large $k$ \cite{Ambainis2012,Sattath2016}; or in its absence, a phase which is unsatisfiable, UNSAT, as happens for $k=2$ \cite{LaumannQSAT2010_2}.
For numerical studies, small $k$ are more tractable than large ones, so that the question of the existence of an ENTSAT phase for $k=3$ is of considerable practical relevance. 

For $k=3$-QSAT, satisfying product states are known to persist up to the threshold $\alpha_{ps} \approx 0.91$, where the core of the interaction graph has density $\beta_c = M_{c}/N_{c} = 1$~\cite{LaumannQSAT2010}. 
Previous numerical diagonalization studies have reported that the SAT phase might persist slightly further to a transition at $\alpha \approx 1 \pm 0.06$ (quoted error bars)~\cite{LaumannQSAT2010}. 

Here, we report a more focused study which reinforces the possibility that the UNSAT transition actually coincides with the disappearance of product states. 
More precisely, our numerical results below suggest an upper bound on the UNSAT transition $\beta \le \beta_c( 1 + 1/17)$, which corresponds to $\alpha \le 0.97$, while we were not able to resolve a stable ENTSAT regime intermediate between the PRODSAT and UNSAT ones. 

The exponential dependence of Hilbert space dimension on size $N$ remains a significant barrier to finite-size exact diagonalization studies---even with Moore's law, this only leads to a linear increase in tractable system size with (human rather than CPU) time. 

In order to make progress, we exploit monotonicity, which states that a given QSAT instance is UNSAT if any subgraph is UNSAT. This holds because adding a projector increases the number of constraints to be satisfied, and so cannot heal the violation of any existing projectors.

Since we are interested in separating $\alpha_c$ from $\alpha_{ps}$, we focus on random instances whose cores are just barely overconstrained as far as product states are concerned, ie. $M_c = N_c + 1$. 
These may be generated by repeatedly sampling $(N,M)$ graphs and stripping them to their cores until such instances are found. By monotonicity, reattaching the stripped projectors can only make SAT instances UNSAT, and not vice versa. 

We use a sparse diagonalization routine on instances of generic projector Hamiltonians in order to determine whether or not these cores are satisfiable---by geometrisation, the outcome does not depend on the realisation of the generic projectors with high probability, so that our averaging focuses on generating different random cores. 

\begin{figure}[tb]
	\centering
	\includegraphics[]{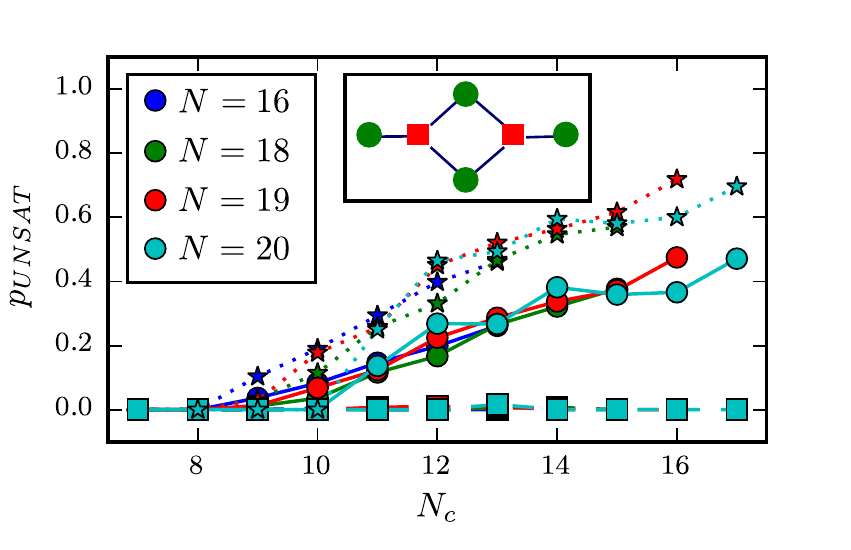}
	\caption{Probability that a random $k=3$-core with $M_c = N_c+1$ is UNSAT. 
	Circles indicate probability across all decidable generated instances; 
	instances with minifans (squares) are essentially always satisfiable at these sizes so that those without (stars) contain all of the UNSAT instances.
	(inset) A minifan motif: two projectors (red squares) share two qubits (green circles).
	}
	\label{fig:unsat-cores-minifans}
\end{figure}

The resulting data for $N = 10-25$, $N_c = 8-17$, with $N_{samp} = 50-500$ samples per $(N, N_c)$ are shown in Fig.~\ref{fig:unsat-cores-minifans}. 
Note that, as $N_c$ increases, we find that the probability of the core being UNSAT increases monotonically. There is no additional dependence on $N$ within statistical fluctuations. 
This is particularly striking as the effective density of the core $\beta = 1 + 1/N_c$ is monotonically \emph{decreasing} with $N_c$, so that upon increasing $N$, one moves closer to the critical value $\beta_c=1$. 

Although for the systems sizes available, $N_c\leq17$, the fraction of UNSAT instances never reaches above $p_{UNSAT}\approx0.5$, there is a considerable finite-size dependence of $p_{UNSAT}$ which shows no clear sign of asymptotically reaching a plateau for these system sizes.

Further evidence that the core goes UNSAT at $\beta = 1$, is provided by the observation that a large fraction of the SAT instances owe their existence to the presence of a geometrical motif in the random graph which we call a `minifan', 
a subgraph of $G$ in which a pair of clauses shares two qubits (see Fig.~\ref{fig:unsat-cores-minifans}, inset). 

The density of minifans in the random graph vanishes with increasing $N$ as $\propto 1/N$  and thus disappears in the thermodynamic limit.
Nonetheless, at accessible sizes, minifans are still quite common and even at $N_c = 17$ roughly $30\%$ of the generated cores have at least one minifan as a subgraph. 
Crucially, at these sizes, when a core contains a minifan, it is satisfiable (squares, Fig.~\ref{fig:unsat-cores-minifans}) and thus filtering them out to better approximate the behavior of the large $N$ ensemble raises the probability of UNSAT significantly (stars) and by itself reduces the fraction of SAT instances by about two fifths for the instances with the largest $N$. 
It seems reasonable to expect that identifying and culling more complex finite-size motifs will further lead to an enhancement of $p_{UNSAT}$. 

Taken together, this does leave a tiny sliver in clause density $\alpha$ for the possible existence of an ENTSAT phase at $k=3$. However, these results provide little evidence of an ENTSAT phase for the systems sizes studied, while being fully consistent with an UNSAT phase setting in at the critical value quoted above.

\section{Entropy of clusters} 
\label{sec:counting_dimer_covering_states}

In this section, we estimate the entropy of dimer coverings $\Sc$ on the core of a random interaction graph $G$, as this provides an estimate of the entropy of reference product states on the core. 
The core is itself a random interaction graph with degree distribution given by a truncated Poisson distribution with mean $\lambda(\alpha)$ (see Appendix~\ref{sec:core_data} for more details) \cite{Mezard2003}.

\subsection{Pauling Estimate} 
\label{sub:pauling_estimate}

Let us construct a simple Pauling-type estimate of $\Sc$ on a core with $N_c$ nodes of fluctuating degrees $d_i$, $i = 1 \cdots N_c$, $M_c$ $k$-clauses, and an appropriate core clause density, $\beta=M_c/N_c$. 
There are a total of $kM_c$ links which can be either occupied or unoccupied. Since each clause must be covered by one dimer, we have $k$ allowed configurations out of the $2^k$ configurations at each clause and $d_i + 1$ allowed configurations out of $2^{d_i}$ configurations at each node. Thus, the total number of allowed configurations can be estimated 
\begin{align}
\# &\approx 2^{kM_c} \left( \frac{k}{2^k}\right)^{M_c} \prod_{i} \left( \frac{1+d_i}{2^{d_i}}\right)\\
&\approx k^{M_c} \prod_{i} \left( \frac{1+d_i}{2^{d_i}}\right)
\end{align}
Taking a logarithm and using $k M_c = k \beta N_c = \sum_i d_i$,
\begin{align}
	\frac{\Sc}{N_c} &= \beta \log k + \frac{1}{N_c}\sum_i \log \left(\frac{1 + d_i}{2^{d_i}}\right)
\end{align}

When $\beta > 1$, we no longer have any product state solutions since there are more nodes than clauses and hence no more dimer coverings. 
Plugging in numbers, the entropy of product state solutions at the transition point for $k=3$ (Appendix~\ref{sec:core_data}) and $\beta = 1$ is $\Sc/N_c \approx 0.37$.




\begin{figure}[t]
  \includegraphics[width=0.7\columnwidth]{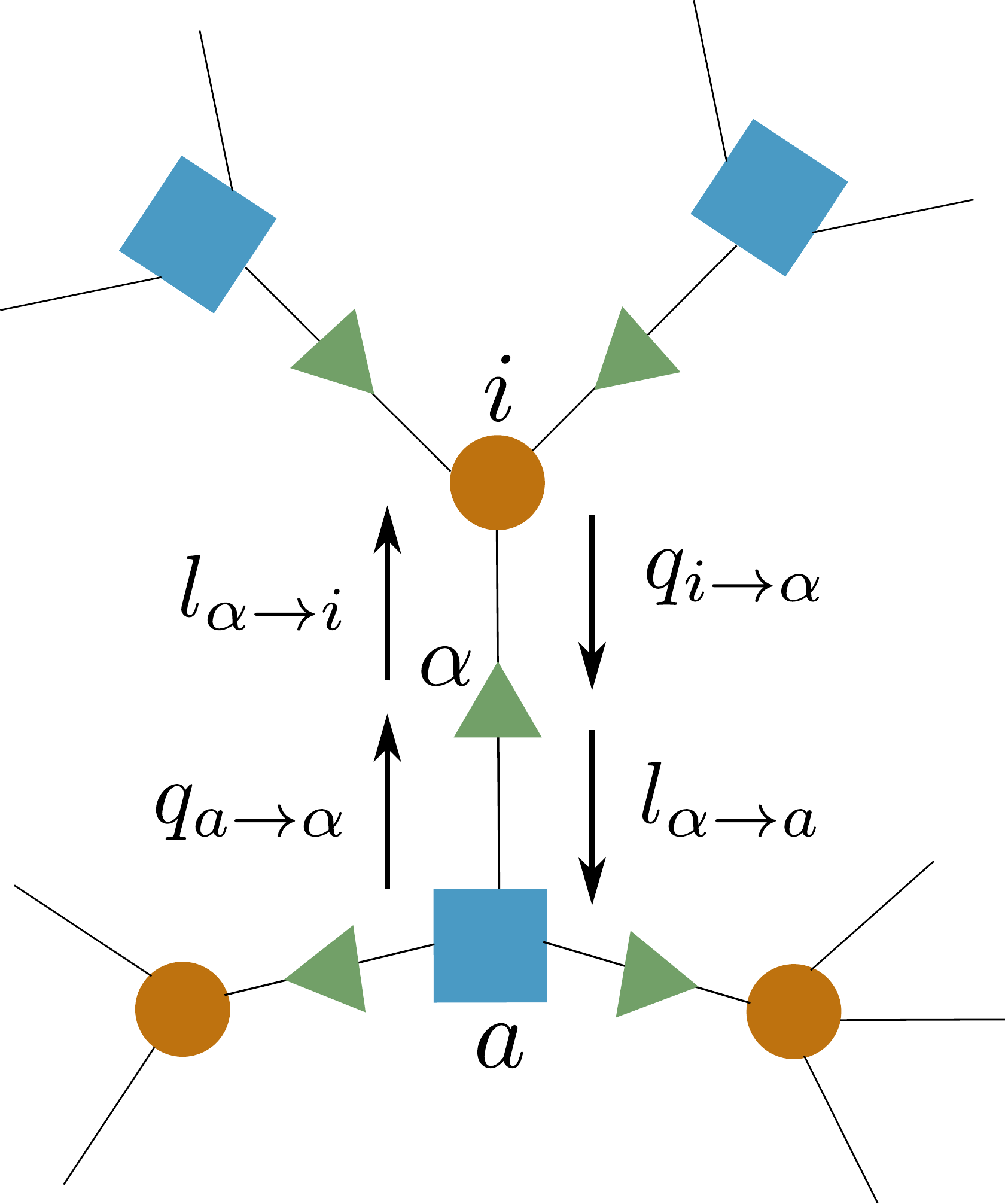}
  \caption{\label{fig:cavity} The cavity method introduces messages between elements of the graph which correspond to probabilities. We introduce additional degrees of freedom on the bonds (triangles) which indicate whether the bond is covered by a dimer or not.}
\end{figure}

\subsection{Cavity Calculation} 
\label{sub:cavity_calculation}
As the core is itself a random interaction graph, its local structure is tree-like and we use the cavity method to count the number of dimer coverings. 
The following treatment is similar to the counting of matchings in random graphs undertaken in \cite{Zdeborova:2006aa}, suitably generalized to the interaction graph structure of $k$-QSAT.
Recall that in a valid dimer covering, we require that every constraint $a$ is covered but not every spin $i$
needs to be covered. 
We define variables $S_{\alpha}$ on the bonds $\alpha$  which denote whether a particular bond is covered by a dimer ($S_{\alpha}=1$) or not ($S_{\alpha}=0$),
and also a fugacity $\lambda$ to the dimer (Fig.~\ref{fig:cavity}(b)). 
Hence we associate a partition function to the system through
\begin{equation}
\mathcal{Z}=\sum_{\left\{ S_{\alpha}\right\} }\left[\prod_{a,i}\mathbb{I}\left(\sum_{\beta\in\partial a}S_{\beta}\leq1\right)\mathbb{I}\left(\sum_{\beta\in\partial i}S_{\beta}\leq1\right)\right]\lambda^{\sum_{\alpha}S_{\alpha}}
\end{equation}
\\
where the term inside the $\left[\ \right]$ enforces the hard-core constraint on dimers. 
In the limit of large $\lambda$, the partition function is dominated by terms corresponding to valid dimer
coverings provided they exist. 

We evaluate the free energy corresponding to this partition function through 
the cavity method since the interaction graph is locally tree-like near the transition. 
We define the appropriate cavity probabilities

\begin{widetext}

\begin{align}
\mathrm{P}_{i\rightarrow\alpha}^{'}\left[S_{\alpha}\right] & ={\displaystyle \sum_{\substack{\left\{ S_{\beta}\right\} ,\\
\beta\in\partial i\backslash\alpha
}
}\mathbb{I}\left(\sum_{\gamma\in\partial i}S_{\gamma}\leq1\right)\lambda^{S_{\alpha}}{\displaystyle \prod_{\beta}\mathrm{P}_{\beta\rightarrow i}\left[S_{\beta}\right]}}\\
\mathrm{P}_{\alpha\rightarrow i}^{'}\left[S_{\alpha}\right] & =\mathrm{P}_{a\rightarrow\alpha}\left[S_{\alpha}\right]\\
\mathrm{P}_{a\rightarrow\alpha}^{'}\left[S_{\alpha}\right] & =\sum_{\substack{\left\{ S_{\beta}\right\} ,\\
\beta\in\partial a\backslash\alpha
}
}\mathbb{I}\left(\sum_{\gamma\in\partial a}S_{\gamma}\leq1\right)\lambda^{S_{\alpha}}{\displaystyle \prod_{\beta}\mathrm{P}_{\beta\rightarrow i}\left[S_{\beta}\right]}\\
\mathrm{P}_{\alpha\rightarrow a}^{'}\left[S_{\alpha}\right] & =\mathrm{P}_{i\rightarrow\alpha}\left[S_{\alpha}\right]
\end{align}
After normalizing, we define the messages (Fig.~\ref{fig:cavity}(b)) 

\begin{align}
q_{i\rightarrow\alpha} & \coloneqq\mathrm{P_{i\rightarrow\alpha}\left[1\right]=}\dfrac{\lambda}{1+\lambda+{\displaystyle \sum_{\beta\in\partial i\backslash\alpha}\frac{l_{\beta\rightarrow i}}{1-l_{\beta\rightarrow i}}}}\\
l_{\alpha\rightarrow i} & \coloneqq\mathrm{P}_{a\rightarrow\alpha}\left[1\right]=q_{a\rightarrow\alpha}\\
q_{a\rightarrow\alpha} & \coloneqq\mathrm{P}_{a\rightarrow\alpha}\left[1\right]=\dfrac{\lambda}{1+\lambda+{\displaystyle \sum_{\beta\in\partial a\backslash\alpha}\frac{l_{\beta\rightarrow a}}{1-l_{\beta\rightarrow a}}}}\\
l_{\alpha\rightarrow a} & \coloneqq\mathrm{P}_{\alpha\rightarrow a}\left[1\right]=q_{i\rightarrow\alpha}
\end{align}

Here, $q_{i\rightarrow\alpha}$ and $l_{\alpha\rightarrow a}$ denote the probability for bond $\alpha$ to be occupied ($=1$) in the absence of $a$; and the other two messages $l_{\alpha\rightarrow i}$ and $q_{a\rightarrow\alpha}$ denote the probability for $\alpha=1$ in the absence of $i$. Although two of the messages are redundant, we have retained them as the resulting equations are more symmetric.

The free energy
of the system can be expressed in terms these messages and consists
of five contributions coming from the three kinds of vertices ($a=\square,i=\bigcirc,\alpha=\triangle$)
and the two types of bonds ($i\alpha=\bigcirc-\triangle,a\alpha=\square-\triangle$).
\begin{align}
\mathrm{F_{a}} & =\log\left[\prod_{\beta\in\partial a}\left(1-l_{\beta\rightarrow a}\right)+\sum_{\beta\in\partial a}\left(l_{\beta\rightarrow a}\prod_{\gamma\in\partial a\backslash\beta}\left(1-l_{\gamma\rightarrow a}\right)\right)\right]\\
\mathrm{F_{i}} & =\log\left[\prod_{\beta\in\partial i}\left(1-l_{\beta\rightarrow i}\right)+\sum_{\beta\in\partial i}\left(l_{\beta\rightarrow i}\prod_{\gamma\in\partial i\backslash\beta}\left(1-l_{\gamma\rightarrow i}\right)\right)\right]\\
\mathrm{F_{\alpha}} & =\log\left[\left(1-q_{a\rightarrow\alpha}\right)\left(1-q_{i\rightarrow\alpha}\right)+\frac{q_{a\rightarrow\alpha}q_{i\rightarrow\alpha}}{\lambda}\right]\\
\mathrm{F_{i\alpha}} & =\log\left[\left(1-l_{\alpha\rightarrow i}\right)\left(1-q_{i\rightarrow\alpha}\right)+\frac{l_{\alpha\rightarrow i}q_{i\rightarrow\alpha}}{\lambda}\right]\\
\mathrm{F_{a\alpha}} & =\log\left[\left(1-l_{\alpha\rightarrow a}\right)\left(1-q_{a\rightarrow\alpha}\right)+\frac{l_{\alpha\rightarrow a}q_{a\rightarrow\alpha}}{\lambda}\right]
\end{align}

The total free energy is then given by 
\begin{align}
\mathrm{F} & =\sum_{a}\mathrm{F}_{a}+\sum_{i}\mathrm{F_{i}}+\sum_{\alpha}\mathrm{F_{\alpha}}-\sum_{i\alpha}\mathrm{F_{i\alpha}}-\sum_{a\alpha}\mathrm{F_{a\alpha}}
\end{align}
We can also express the average occupation of a constraint $a$ by
a dimer ($\langle n_{a}\rangle\coloneqq N_{\mathrm{dimer}}^{c}/M^{c}$)
through 
\begin{align}
\langle n_{a}\rangle & =\dfrac{{\displaystyle \sum_{\beta\in\partial a}}l_{\beta\rightarrow a}{\displaystyle \prod_{\gamma\in\partial a\backslash\beta}}\left(1-l_{\gamma\rightarrow a}\right)}{{\displaystyle \sum_{\beta\in\partial a}}l_{\beta\rightarrow a}{\displaystyle \prod_{\gamma\in\partial a\backslash\beta}}\left(1-l_{\gamma\rightarrow a}\right)+{\displaystyle \prod_{\beta\in\partial a}}\left(1-l_{\beta\rightarrow a}\right)}
\end{align}
and hence we can obtain the entropy density of dimer coverings at
a given clause density on the core $\beta$ as
\begin{align}
\frac{\Sc}{N^{c}} & =\frac{F}{N^{c}}-\frac{N_{dimer}^{c}}{N^{c}}\log\lambda \nonumber\\
 & =\frac{F}{N^{c}}-\beta\langle n_{a}\rangle\log\lambda
\end{align}

For regular random graphs with $k = d = 3$ (corresponding to $\beta = 1$), we can solve the equation exactly to obtain
\begin{align}
\langle n_{a}\rangle & \approx1-0.47\lambda^{-0.5}+0.06\lambda^{-1}+\mathcal{O}\left(\lambda^{-1.5}\right) \\
\dfrac{\Sc}{N^{c}} & \approx0.29+0.94\lambda^{-0.5}-0.06\lambda^{-1}+\mathcal{O}\left(\lambda^{-1.5}\right)
\end{align}

\end{widetext}

\begin{figure}
\includegraphics[width=\columnwidth]{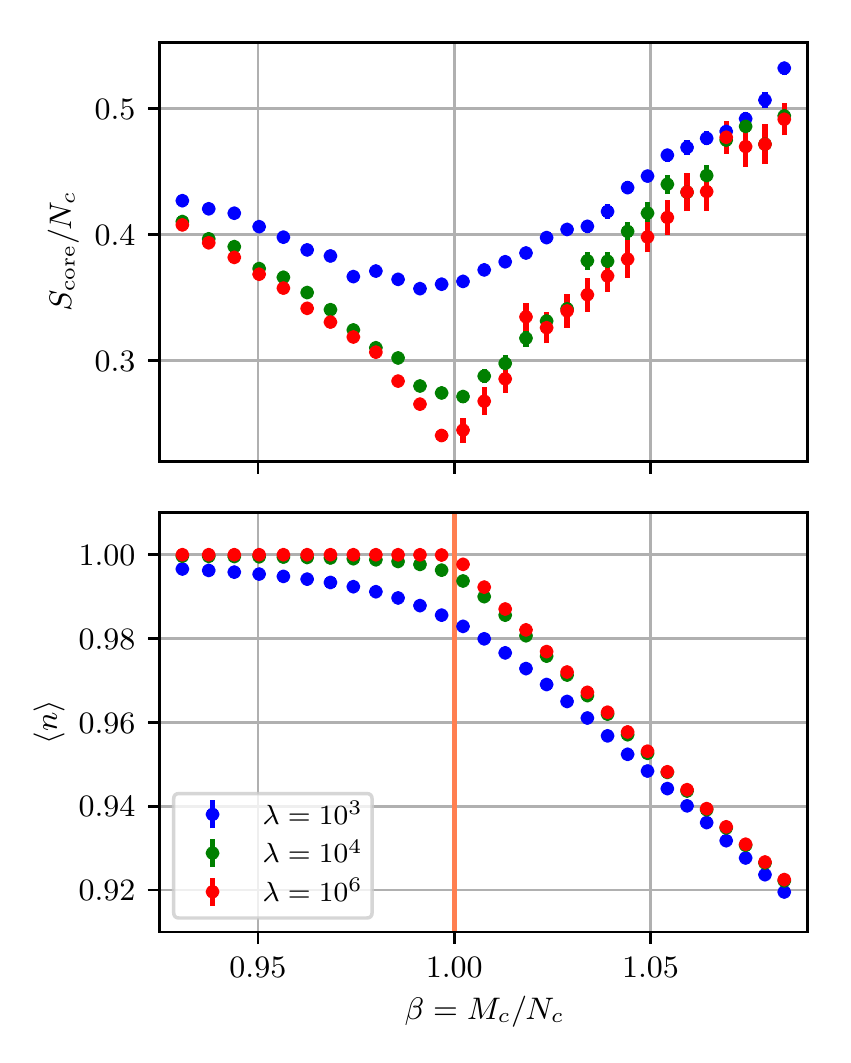}
	\caption{\label{fig:cavity_result} 
    Cavity estimate of core entropy per core spin variable (top) and average occupation of constraints (bottom) as a function of core density $\beta$ at various fugacities $\lambda$. 
}
\end{figure}

For Erdos-R\'enyi random graphs, we use population dynamics to find a fixed point solution to the cavity equations. We initialize a population of $10^4$ random messages and iterate the cavity equations for $4000$ steps after which convergence is achieved and we obtain a fixed-point distribution. 
The free energy is then sampled from this fixed point distribution to obtain the entropy density $\Sc/N_c$ and average occupation $\langle n_a \rangle$. The entropy density corresponding to exact dimer coverings is obtained in the limit of large fugacity $\lambda$.

In principle, one can bypass the fugacity and instead directly impose a hard constraint in the partition function that every interaction is covered by a dimer. 
This treatment leads to instabilities in the cavity equations and the associated population dynamics. 
This is unsurprising as the partition function itself is exponentially large at the critical point $\beta = 1$ and formally zero beyond it since there are no coverings. 
The introduction of a finite fugacity allows the system to find the maximal covering that fits even for $\beta > 1$ and accordingly undergo a simple first order transition at the critical point.
Analogous instabilities arise in the treatment of the matching problem on random graphs \cite{Zdeborova:2006aa}.

\begin{figure}
 \includegraphics[width=\columnwidth]{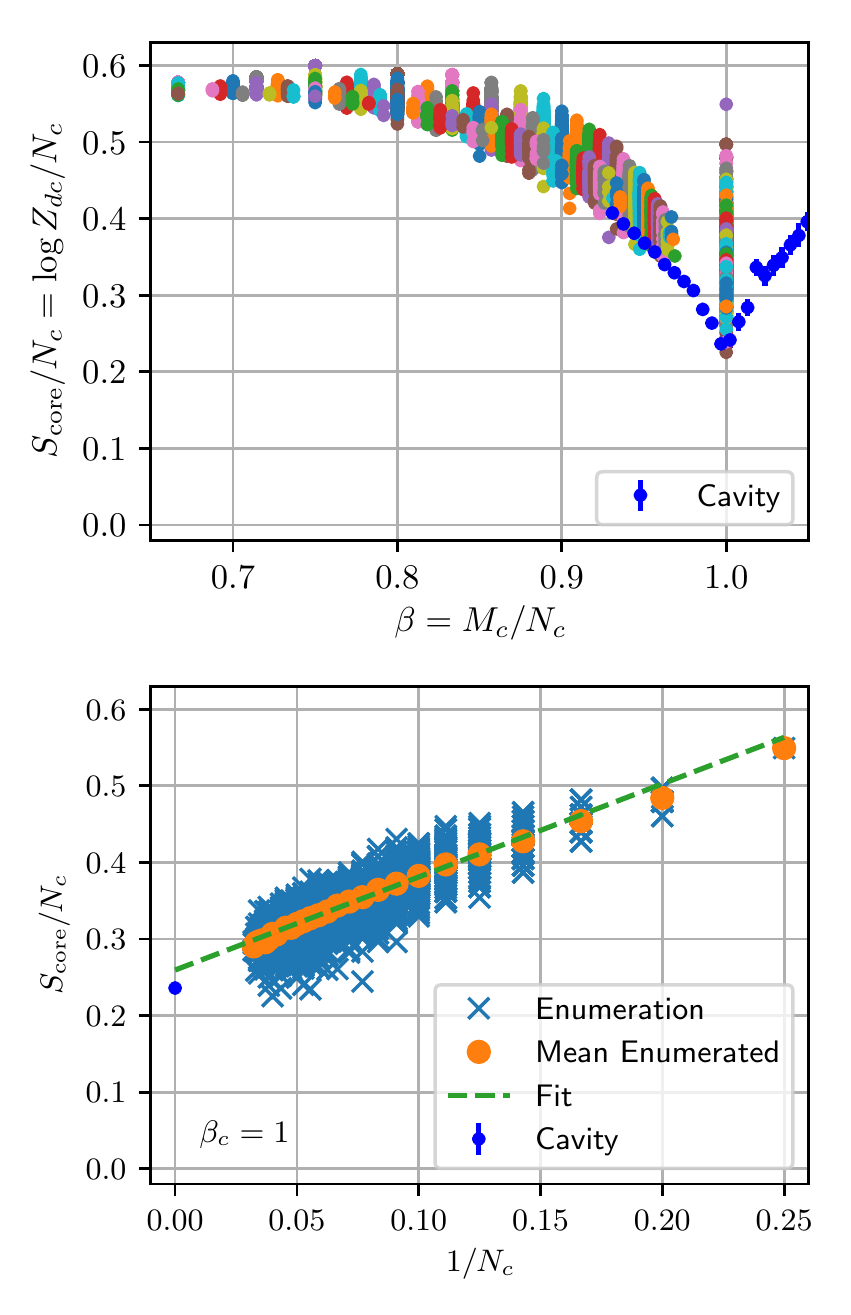}
	\caption{\label{fig:results} Entropy of dimer coverings of the core as a function of core density $\beta$ (top) and at the critical point ($\beta = \beta_c = 1$) as a function of core size $N_c$ (bottom). 
    (top) The different color markers correspond to instances from cores of sizes up to $N_c = 35$. At a given density, larger sizes flow to smaller $\Sc$, although this is hard to see in the figure. The cavity estimate is the value extrapolated in the limit of infinite fugacity $\lambda$.
    (bottom) The finite size scaling of $\Sc/N_c$ at the critical point is consistent with a fit linear in $1/N_c$. 
 }
\end{figure}

Finally, for small system sizes, we also use an exhaustive enumeration procedure to find all possible dimer coverings.
We find agreement between the two algorithms and compute $\Sc$ for various values of $\alpha_{c}$ (see Fig.~\ref{fig:results}). 
For $\beta< \beta_c = 1$, the covering problem on the core is under constrained. There are a number of dimer coverings that leave spins uncovered and hence many \emph{small} dimer flips can connect differing possible dimer coverings through monomers representing uncovered qubits. This situation can be contrasted with the one at $\beta=1$ where we have equal numbers of constraints and qubits so that every node in the core is covered by a dimer (Fig~\ref{fig:interactionGraph}). In this fully packed situation, there is still a macroscopic number of macroscopically distinct dimer coverings with an entropy density $S/N_c \approx 0.23$. Any two dimer coverings differ along \emph{large} loops in the graph. Finally, for $\beta>1$, there are more clauses than variables and the entropy corresponds to dimer coverings which leave some clauses uncovered and result in a density of monomers. In this situation there are no valid dimer coverings of the graph. (We note in passing that while there are no dimer coverings beyond this point, it has been found that product states can still give good approximations to the ground states of QSAT.~\cite{Hsu2013})



\section{Internal entropy of a cluster} 
\label{sub:relation_to_qsat}
Now that we have estimated the entropy due to the macroscopically distinct product states on the core, we can construct the full state by reattaching the hair and satisfying the projectors on it by constructing dimer coverings on the hair (Eq.~\ref{eq:DCHair}). 
Unlike dimer coverings on the core, there are free spins on the hair which allow local rearrangements (Fig.~\ref{fig:interactionGraph}).
However, these free spins do not allow rearrangement on the core since the core is fully packed. 
The total zero energy entropy at the critical point is thus given by sum of the entropy of clusters and the internal entropy of each cluster due to degeneracy associated to the hair,
\begin{equation}
S_{\mathrm{total}} = \Sc + S_{\mathrm{hair}}
\end{equation}
For a given dimer covering on the core, the hair entropy has two components -- a discrete geometric component due to rearrangement of the dimer coverings and a `zero mode' contribution due to the continuous manifold of satisfying product states associated even to a fixed dimer covering due to the free spins. 
A naive estimate of the geometric entropy is given by 
\begin{align}
S_{\mathrm{hair}}^{\mathrm{geometric}} = S_2(1 - (M - M_c)/(N - N_c))
\approx 0.5N_{h}
\end{align}
where $N_h$ is the number of spins on the hair. 

Below, we estimate the zero mode entropy corresponding to a fixed dimer covering on the full graph.

\subsection*{Zero mode entropy}

Restricting the allowed states to be product states in QSAT gives us PSAT \cite{Hsu2013}. Zero modes in the PSAT problem correspond to finite dimensional, zero energy submanifolds of  product state space $(\mathbb{CP}^1)^N$. In the quantum world, arbitrary superpositions of the zero energy product states satisfy the related QSAT problem. Thus, the linear span of the (nonlinear) product state manifold is contained in the satisfying space for QSAT and its dimension provides a lower bound on the QSAT degeneracy of a cluster.

In this section, we provide an entropic estimate for the dimension of the quantum zero energy space due to the presence of PSAT zero modes. To warm up, we consider a connected instance of QSAT whose clauses project onto product states and select a zero energy state associated with a particular dimer covering. In this case, the $N-M$ uncovered spins are completely free while the $M$ covered spins are completely fixed by the dimer covering. The span of this manifold of product states is thus precisely the Hilbert space of $d=N-M$ free qubits, with dimension $2^d$. 

Let us now try to count this $2^d$ dimensional space by local arguments. We take a particular zero energy state $\ket{\Omega}$ and choose local bases such that it corresponds to the all $\up$ state, $\ket{\Omega} = \ket{\up\up\cdots\up}$. Let $\set{w^{\alpha}_i}_{\alpha=1\cdots d}$ be an orthonormal basis for the $d=N-M$ dimensional kernel of the linearized zero energy condition. The satisfying product states in a small neighborhood of $\ket{\Omega}$ are then parameterized by $d$ complex coordinates $\delta c_\alpha$:
\begin{align}
	\ket{\Psi(\delta c_\alpha)} &= \prod_{j=1}^N(\ket{\up} + \delta c_\alpha w^\alpha_j \ket{\down}) \label{eq:prodexpansion}\\
	&=\ket{\Omega} + \delta c_\alpha \sum_j w_j^\alpha b^\dagger_j \ket{\Omega} \nonumber\\
	&+ \f{1}{2} \delta c_\alpha \delta c_\beta P \sum_{j, k}w^\alpha_j w^\beta_k b^\dagger_j b^\dagger_k \ket{\Omega} + \cdots \nonumber
\end{align}
where we have introduced a collection of bosonic spin flip operators $b^\dagger_j = \ket{\down}_j\!\bra{\up}$ and the projector $P$ implements a hard core constraint on the bosons. In this language, each order $n$ in the expansion in $\delta c$ lives in the $n$ boson sector of the full Hilbert space. 

Since each order is algebraically independent in $\delta c$, the span of the product states in the neighborhood of $\ket{\Omega}$ is the direct sum of the span of the $n$-boson states at each order. A generic state at order $n$ can be written
\begin{align}
	\f{1}{n!} \left( \sum_{\delta c} \psi_{\delta c} \delta c_{\alpha_1} \cdots \delta c_{\alpha_n}\right) P w^{\alpha_1}_{j_1} \cdots w^{\alpha_n}_{j_n} b^\dagger_{j_1}\cdots b^\dagger_{j_n} \ket{\Omega}
\end{align}
where the sum runs over an arbitrary finite collection of $\delta c$ vectors.
This expression has two parts: the left hand bracketed expression is a symmetric rank $n$ tensor over $\mathbb{C}^d$ (aka. an element of $\mathrm{Sym}^n \mathbb{C}^d$). This tensor of coefficients is a completely general symmetric tensor since the product tensors of the form $\delta c \otimes \delta c \cdots \otimes \delta c$ span this space 
\footnote{
We wish to show that states of the form $\delta c \otimes \delta c \cdots \otimes \delta c$ span the space $\mathrm{Sym}^n \mathbb{C}^d$. This space
is isomorphic to the space of degree $n$ polynomials in the $d$ variables $e_1,\cdots,e_d$ representing a basis of the space $\mathbb{C}^d$. In particular, this space is spanned by monomials of the form $e_1^{n_1}\cdots e_d^{n_d}$ where $\sum n_i = n$ for nonnegative integers $n_i$. 
If we show that these monomials are in the span of states of the form $f(a_i) = (\sum_i a_i e_i)^n$, then we are done. Observe that derivatives of $f$ with respect to the coefficients $a^i$ are in the span of the image of $f$. But $\f{\partial^n f}{\partial^{n_1} a_1 \partial^{n_2} a_2\cdots\partial^{n_d}a_d} = n! e_1^{n_1}\cdots e_d^{n_d}$, so we are done.
}.
The second part is a state constructed from $n$ bosons placed into any of $d$ modes with a hard-core constraint imposed by $P$. 

In the special case of product projectors with $d$ free spins, the mode functions $w^\alpha$ are completely localized on lattice sites. By relabeling the sites, we can write this $w^\alpha_j = \delta^\alpha_j$ and then simplify the product state expansion
\begin{align}
	\sum_{\delta c} \psi_{\delta c} \ket{\Psi(\delta c)} &\propto \ket{\Omega} + (\sum_{\delta c} \psi_{\delta c}\delta c_\alpha)  b^\dagger_\alpha \ket{\Omega} \\
	&+ \f{1}{2} (\sum_{\delta c} \psi_{\delta c} \delta c_\alpha \delta c_\beta) P b^\dagger_\alpha b^\dagger_\beta \ket{\Omega} + \cdots
\end{align}
With this expansion, we see that the state space at order $n$ is precisely that of $n$ hardcore bosons on $d$ lattice sites. This has dimension
\begin{align}
	D_n = \binom{d}{n}
\end{align}
and, we find a total dimension
\begin{align}
	D = \sum_n D_n = 2^d
\end{align}
recovering precisely the counting found for $d$ free spins above.

More generally, for generic non-product projectors, we expect the mode matrix  $w^\alpha_j$ to have non-zero entries for all $\alpha$ and $j$ on the hair. The mode matrix might also show some additional structure due to the presence of disconnected clusters of hair, but we do not consider this aspect here. In this case, the dimension of the space spanned at order $n$ is no larger than the space $\mathrm{Sym}^n \mathbb{C}^d$ spanned by the coefficient tensor in $\delta c^\alpha$ and the space of $n$ hardcore bosons in $N_{h}$ sites. The relevant number of sites here is the number of sites on the hair ($N_{h}$) since the core is fully packed and rearrangements on the core and hair don't affect each other.
\begin{align}
	D_n \le \min\Set{\binom{n+d-1}{n}, \binom{N_{h}}{n}}.
\end{align}
Since the binomial coefficients monotonically increase with the top argument, the minimizer is given by $\binom{n+d-1}{n}$ for $n+d-1 \le N_{h}$. 
Thus, the total dimension is 
\begin{align}
	D = \sum_n D_n \le \sum_{n=0}^{N_{h}-d}\binom{n+d-1}{n} + \sum_{n=N_{h}-d+1}^{N_{h}} \binom{N_{h}}{n}
\end{align}
If we now assume that $d = \gamma N_{h}$, we may estimate this dimension to leading exponential order in $N_{h}$ by steepest descent. The first term is
\begin{align}
	\sum_{n=0}^{N-d}\binom{n+d-1}{n} &\sim N_{h} \int_0^{1-\gamma} dx\, \binom{(x+\gamma)N_{h}}{xN_{h}} \nonumber \\
	&\sim N_{h} \int_0^{1-\gamma} dx\, e^{N_{h} f(x)}
\end{align}
where the exponent function follows from Stirling's formula as
\begin{align}
	f(x) = -\left[ x\log\f{x}{x+\gamma} + \gamma \log \f{\gamma}{x+\gamma} \right].
\end{align}
As $f(x)$ is monotonically increasing on the range of the integral, the integral is dominated by the right end point:
\begin{align}
	e^{N_{h} f(1-\gamma)} &= e^{N_{h} S_2(\gamma)}
\end{align}
where $S_2(\gamma)$ is the entropy of a single coin flip with probability $\gamma$.

Similarly, the second term is 
\begin{align}
	\sum_{n=N_{h}-d+1}^N \binom{N_{h}}{n} &\sim N_{h} \int_{1-\gamma}^1 dx\, \binom{N_{h}}{xN_{h}} \nonumber \\
	&\sim N_{h} \int_{1-\gamma}^1 dx\, e^{N_{h} S_2(x)}
\end{align}
The entropy function $S_2$ has a maximum at $x=\f{1}{2}$ corresponding to an entropy $\log(2)$. If $\gamma > \f{1}{2}$, the integral includes this saddle point and the result is $\sim 2^{N_{h}}$ to leading exponential order. If $\gamma < \f{1}{2}$, the integral is dominated by its left end-point and we have
\begin{align}
	N_{h} \int_{1-\gamma}^1 dx\, e^{N_{h} S_2(x)}\sim e^{N_{h} S_2(1-\gamma)} 
\end{align}
Putting this together, we find the QSAT entropy per spin due to the zero modes to be
\begin{align}
	\frac{S_{\mathrm{zero}}}{N_h} = \frac{1}{N_h}\log D \le \left\{ \begin{array}{ll} \log(2) & \gamma > \f{1}{2} \\
	S_2(\gamma) & \gamma < \f{1}{2} \end{array}\right.
\end{align}

The result is non-rigorous as it relies on analyzing only a leading order expansion near a given product state, but it provides a nontrivial conjecture for the dimension of the space of states spanned by the PSAT zero modes. The estimated entropy per spin is significantly greater than the $\log(2) \gamma$ associated with having $d=\gamma N_{h}$ strictly free spins. 


\section{Conclusions and outlook\label{sec:conclusion}}
We have generalized the notion of clustering to quantum systems. 
Unlike the classical definition involving solutions far separated by a Hamming distance, here we define a space as clustered if it can be decomposed into a set of orthogonal subspaces which are macroscopically distinct.
That is, matrix elements of local operators vanish between states in the different subspaces. 
On the other hand, each of these subspaces can be constructed from local operations on a reference short-range correlated state whose connected correlations decay at large distances.
We thus identify each subspace as a cluster in the quantum ground space.

We have provided evidence that there is a direct transition from the PRODSAT to UNSAT phase in 3-QSAT. Moreover, quantum clustering occurs in the ground state space at the critical point separating these two phases.
Our argument is essentially constructive as we can identify reference product states from which the cluster subspaces may be generated.
These reference product states are in one-to-one correspondence with dimer coverings on the core of the interaction graph.

The entropy of clusters is
\begin{align}
	S_{\mathrm{core}} \approx 0.14 N
\end{align}
Each cluster carries an additional internal entropy $S_{hair}$, which arises from two contributions.
First, for a fixed reference dimer covering, there is entropy due to the zero modes
\begin{align}
	S_{\mathrm{zero}} = N_h S_2(0.2) \approx 0.2 N
\end{align}
(using $N_h / N \approx 0.4$).
This is significantly more than would be expected from the naive estimate that each free spin on the hair contributes $\log(2)$ which would give an entropy $\approx 0.06N$.

The second contribution to the internal entropy arises due to monomer rearrangements on the hair. 
While it is possible to estimate this hair entropy from calculations analogous to those presented in Sec.~\ref{sub:cavity_calculation} on the full interaction graph geometry, it is clear that the zero mode entropy and this additional geometric entropy interact.
It would be very interesting to develop a theory of this interaction.
Here, we content ourselves with a simple upper bound for the total entropy due to the hair by considering all of the hair spins to be completely unconstrained
\begin{align}
S_{\mathrm{hair}} < N_h \log (2) \approx 0.28 N
\end{align}

The quantum clustering in this model is particularly striking as it provides a precise quantum parallel between the structure of solution space for 1RSB classical models, such as $k$-XORSAT~\cite{Mezard2003}, and the structure of the zero energy Hilbert space for $k$-QSAT.
More broadly speaking, we have used the discrete nature of the solution space of $k$-QSAT to identify a classical quantity -- dimer coverings of the core of the 
interaction graph -- based on which clustering appears naturally in the space of satisfying solutions embedded in the full Hilbert space. 
It will be interesting to see in what further settings clustering appears. 

In particular, it is intriguing to ask if there exists a more 'intrinsically' quantum mechanical definition of clustering. One natural approach would be to aim for a prescription of equivalence classes of ground states based on their connectedness via a set of natural local operations. 
Given the wide range of possible `landscapes' in -- and the continuous nature of  -- Hilbert space, this may very well turn up much richer structures than even the hierarchical clustering discussed for the solution spaces of classical optimisation problems \cite{SpinGlassTheoryBook1986}.

\begin{acknowledgments}
CRL acknowledges the hospitality of the International Center for Theoretical Physics, Trieste, the Max-Planck Institute for the Physics of Complex Systems, Dresden, the Aspen Center for Physics, the Kavli Institute for Theoretical Physics and the support of the Sloan Foundation through a Sloan Research Fellowship and the NSF through Grant No. PHY-1656234. 
Note that any opinion, findings, and conclusions or recommendations expressed in this material are those of the authors and do not necessarily reflect the views of the National Science Foundation. SLS acknowledges the support of the John Templeton Foundation. The work in Dresden was supported by the German Science Foundation (DFG) via grant SFB 1143. 
\end{acknowledgments}

\bibliography{refs}

\appendix

\section{Core data} 
\label{sec:core_data}

We summarize a few results derived in~\cite{Mezard2003} regarding the core of random interaction graphs $G$. The core is the maximal subgraph of $G$ on which all qubit nodes have degree at least 2. The quoted results follow from analyzing the leaf removal algorithm applied to the random interaction graph. 
In the notation of~\cite{Mezard2003}, the clause density $\alpha$ is $\gamma$ and the number of spins per clause $k$ is $p$.

The degree distribution (per node of $G$) on the core is Poissonian for degrees $d\ge 2$:
\begin{align}
	P_c(d) = \left\{ \begin{array}{ll} 0 & \textrm{for } d=0,1\\
	e^{-\lambda^*(\alpha)}\f{\lambda^*(\alpha)^d}{d!} & \textrm{for } d\ge 2 \end{array} \right.
\end{align}
where the parameter $\lambda^*$ corresponds to the largest solution of the equation:
\begin{align}
	e^{-\lambda^*}-1+\left(\f{\lambda^*}{k\alpha}\right)^{\f{1}{k-1}} = 0.
\end{align}

The total number of nodes in the core is
\begin{align}
	N_c(\alpha) = N \sum_{d \ge 2} P_c(d) = N\left[ 1-(1+\lambda^*)e^{-\lambda^*}\right]
\end{align}
and the total number of clauses
\begin{align}
	M_c(\alpha) = N\f{\lambda^*}{k}(1-e^{-\lambda^*}).
\end{align}
For large $\alpha$, the core takes over most of the graph up to exponentially small corrections: $\lambda^* \approx k \alpha$ and $N_c = N(1- (1+k \alpha)e^{- k \alpha})$, $M_c = N \alpha(1-e^{-k\alpha})$.

For reference, at the critical point for $k=3$,
\begin{align}
	\lambda^*(\alpha_c \approx 0.917) &\approx 2.149
\end{align}


\end{document}